\pdfoutput=1
\documentclass[letterpaper]{article}
\usepackage{aaai}
\usepackage{times}
\usepackage{helvet}
\usepackage{courier}
\usepackage{graphics}
\usepackage{graphicx}
\usepackage{amsmath}
\usepackage{amssymb}

\newcommand{\rar}{\rightarrow}
\newcommand{\ra}{\rangle}
\newcommand{\la}{\langle}
\newcommand{\ttt}{\texttt}
\newcommand{\mbb}{\mathbb}
\newcommand{\mbf}{\mathbf}
\newcommand{\mca}{\mathcal}

\begin{document}
\title{An Evidential Path Logic for Multi-Relational Networks}
\author{Marko A. Rodriguez \\
Center for Non-Linear Studies \\
Los Alamos National Laboratory \\
Los Alamos, New Mexico 87545 \\
\And
Joe Geldart \\
Department of Computer Science \\
University of Durham \\
South Road, Durham, DH1 3LE \\
}
\maketitle
\begin{abstract}
\begin{quote}
Multi-relational networks are used extensively to structure knowledge. Perhaps the most popular instance, due to the widespread adoption of the Semantic Web, is the Resource Description Framework (RDF). One of the primary purposes of a knowledge network is to reason; that is, to alter the topology of the network according to an algorithm that uses the existing topological structure as its input. There exist many such reasoning algorithms. With respect to the Semantic Web, the bivalent, monotonic reasoners of the RDF Schema (RDFS) and the Web Ontology Language (OWL) are the most prevalent. However, nothing prevents other forms of reasoning from existing in the Semantic Web. This article presents a non-bivalent, non-monotonic, evidential logic and reasoner that is an algebraic ring over a multi-relational network equipped with two binary operations that can be composed to execute various forms of inference. Given its multi-relational grounding, it is possible to use the presented evidential framework as another method for structuring knowledge and reasoning in the Semantic Web. The benefits of this framework are that it works with arbitrary, partial, and contradictory knowledge while, at the same time, it supports a tractable approximate reasoning process.
\end{quote}
\end{abstract}

\noindent Knowledge structures are used to represent facts about the world. The most common formal data structure to represent knowledge is the network. With respect to symbolic knowledge representation, the multi-relational network (also known as an edge labeled graph or semantic network) is widely used. A multi-relational network is composed of a set of vertices and a family of edge sets, where each edge set has a different nominal, or categorical, label. Formally, a multi-relational network can be represented as $M = (V, \mbb{E})$, where $V$ is the set of vertices and $\mbb{E} = \{E_0, E_1, \dots, E_m \subseteq (V \times V)\}$ is the family of directed edge sets. In recent years, perhaps the most popular instance of a multi-relational data structure for knowledge representation is the Resource Description Framework \cite{dau:rdfgraph2006} of the Semantic Web initiative \cite{lee:semantic2001}\footnote{Other formal models of RDF include a bipartite graph \cite{hayes:birdf2004} and hypergraph \cite{hyperrdf:2006} representation.}. An edge in an RDF network is called a statement, or triple, as it is composed of a subject, predicate, and object. For example, suppose the statement $(i,k,j)$. This statement denotes that $i$ is related to $j$ by means of an $k$-type relationship. Given the previous definition of $M$, this is equivalent to the directed edge $(i,j) \in E_k$. A particular instance of a statement is $(\small{\ttt{marko}}, \small{\ttt{coauthor}}, \small{\ttt{joe}})$. This statement denotes that Marko has a coauthorship relationship to Joe. Languages such as the RDF Schema (RDFS) and the Web Ontology Language (OWL) impose a set of constructs that serve to structure knowledge in a particular manner. The particularities of such a structure are used by an RDFS or OWL reasoner to infer new statements that can be added to the RDF network. The statements inferred by such reasoners are bivalent in that they are either true or false and moreover, their truth value is monotonic as it does not change once it has been asserted.

While RDFS and OWL are common languages in the Semantic Web, the flexibility of RDF can easily support other knowledge structures and reasoning algorithms. The purpose of this article is to present a non-bivalent, non-monotonic, evidential logic and reasoner for multi-relational networks that leverages many of the ideas from Non-Axiomatic Logic (NAL) \cite{nal:wang2006} and the Non-Axiomatic Reasoning System (NARS) \cite{nars2.2:wang}. The philosophical foundation of an evidential logic is that no statement is inherently true or false and that a statement only maintains levels of evidence to support or negate its claim. The notion of
\begin{footnotesize}
\begin{quote}
\textit{experience-grounded semantics} [is where] the truth value of a judgment indicates the degree to which the judgment is supported by the system's experience. Defined in this way, truth value is system-dependent and time-dependent. Different systems may have conflicting opinions, due to their different experiences. \cite{inherit:wang1994}
\end{quote}
\end{footnotesize}
The typical metaphor in an evidential logic system is that of an agent that perceives the world, represents its perceptions in an internal knowledge structure, and reasons on that structure to infer new knowledge \cite{coglogic:wang2004}. Moreover, it is assumed that this agent has limited computational resources in terms of both space and time and thus, does not maintain an objective knowledge structure nor does it necessarily have the ability to reason across its entire subjective knowledge structure. In other words, the agent has only so much information that it can store and process at any one time. This notion is known as the Assumption of Insufficient Knowledge and Insufficient Resources (AIKIR). Non-axiomatic logic is contrasted to axiomatic logic, where truth is bivalent, is defined independent of the time and space requirements necessary to derive it, can be reasoned from a finite set of premises, and where all reasoning produces true, immutable conclusions.

The evidential logic presented in this article forms an algebraic ring over a multi-relational network (i.e.~the knowledge structure) equipped with two binary operations (i.e.~the atoms of the inferencing algorithms). Given the logic's multi-relational formulation, it is possible to comfortably represent this structure in RDF and thus, on the Semantic Web. The primary contribution of this article is the application of evidential logics to multi-relational networks and the formulation of an algebraic evidential reasoner.

\section{Evidence in an Inheritance Network}

With evidential logics, there does not exist an objective boolean truth value for every question that can be asked of the world as the world is not reasoned on directly \cite{sem:wang2004}. What is reasoned on is the agent's internal knowledge structure. For the agent, knowledge is gained as new evidence from the world is discovered (either through direct perception or through communication) or as knowledge is inferred given the agent's internal reasoning system. The Non-Axiomatic Reasoning System (NARS) is an example of an evidential reasoning system \cite{nars2.2:wang}. The data structure proposed for NARS version 2.2 is a directed evidence network denoted $G = (V,E,\lambda)$, where $V$ is a set of vertices, $E \subseteq (V \times V)$ is a set of directed ``inheritance" edges, and $\lambda: E \rar \la [0,1], [0,1] \ra$ maps each edge to an evidence tuple. For example, an edge is denoted
\begin{equation*}
	(i, \la w^+,w^- \ra, j),
\end{equation*}
where $i$ is the tail of the inheritance edge, $j$ is the head of the inheritance edge, and $\la w^+, w^- \ra$ is the evidence tuple for that edge. Moreover, $\la w^+, w^- \ra \equiv \lambda(i,j)$. The meaning of $(i, \la w^+,w^- \ra, j)$ is that there is $w^+$ positive evidence supporting the claim that $i$ $\small{\ttt{isA}}$ $j$ and $w^-$ negative evidence that $i$ $\small{\ttt{isA}}$ $j$. Another interpretation of this edge is that, upon examination, it appears, to the agent, that $i$ has $w^+$ properties in common with $j$ and $w^-$ properties not in common with $j$. This idea is also expressed as $i$ being the intension of $j$ and $j$ being the extension of $i$ \cite{inherit:wang1994}.

In NARS, the evidence tuple of an edge is revised by means of external perception or internal reasoning. The act of perceiving $i$ and $j$ in the external world will return evidence of their relationship and thus, increment $w^+$ or $w^-$ accordingly. With respect to internal reasoning, there are four syllogisms that can be applied to the agent's internal knowledge network: deduction, induction, abduction \cite{syllogism:patzig1968}, and exemplification. For deduction,
\begin{equation*}
	(i, \la w^+_1,w^-_1 \ra, j), (j, \la w^+_2,w^-_2 \ra, k) \rar (i,  \la w^+_3,w^-_3 \ra, k).
\end{equation*}
For induction,
\begin{equation*}
	(i, \la w^+_1,w^-_1 \ra, j), (i, \la w^+_2,w^-_2 \ra, k) \rar (j,  \la w^+_3,w^-_3 \ra, k).
\end{equation*}
For abduction,
\begin{equation*}
	(i, \la w^+_1,w^-_1 \ra, j), (k, \la w^+_2,w^-_2 \ra, j) \rar (i,  \la w^+_3,w^-_3 \ra, k).
\end{equation*}
Finally, there is a less widely used fourth syllogism known as exemplification \cite{histlogic:boche1970,inherit:wang1994}. Exemplification is defined as
\begin{equation*}
	(i, \la w^+_1,w^-_1 \ra, j), (j, \la w^+_2,w^-_2 \ra, k) \rar (k,  \la w^+_3,w^-_3 \ra, i).
\end{equation*}
The values for $\la w^+_3, w^-_3 \ra$ depend upon the specific inference rules of the the evidential reasoner. In \cite{nal:wang2006}, it is explicitly stated that the rules presented are not set in stone, but rather subject to revision themselves as more is understood about the design of evidential systems.

The evidence tuple $\la w^+, w^- \ra$ of an edge can be transformed into a normalized ``truth value" consisting of a new tuple $\la f, c \ra \in \la \emptyset \cup [0,1], \emptyset \cup [0,1] \ra$. The first component $f$ is the frequency of positive evidence and is defined as
\begin{equation*}
	f = \frac{w^+}{w^+ + w^-}.
\end{equation*}
The second component $c$ is the confidence in the stability of $f$ as $k$-more observations are made and is defined as
\begin{equation*}
c = \frac{w^+ + w^-}{(w^+ + w^- + k)}.
\end{equation*}
The parameter $k \in \mbb{R}_0^+$ is a user-defined, system constant. In words, as more positive evidence accumulates relative to negative evidence, $f$ increases towards $1$. As more total evidence accumulates relative to some constant $k$, $c$ increases towards $1$. If there is no evidence, then the edge has an $f$-component of $\emptyset$ which means that the relationship is unknown. Finally, hard ``truth" can be modeled with an evidence tuple of $\la 1, 0 \ra$ with $k = 0$ and thus, an $fc$-tuple of $\la 1, 1 \ra$.

The contribution of this article is to extend the aforementioned evidential logic framework to multi-relational networks composed of both inheritance and non-inheritance edges. Moreover, this article contributes an algebraic ring formulation of evidential reasoning which situates the reasoner within well understood mathematics. From this multi-relational foundation, evidential reasoning can be comfortably executed in the RDF-rich world of the Semantic Web.

\section{Evidential Reasoning using Path Expressions}

A path algebra to map a multi-relational network to a single-relational form was originally presented in \cite{pathalg:rodriguez2008}. The motivation behind the algebra was to provide a formal means by which the large class of single-relational network analysis algorithms could be applied to multi-relational networks in a meaningful way. With respect to this article, the binary operations of $+$ and $\cdot$ are updated so as to work with evidence tuples. 

A multi-relational, evidence network is defined as $M = (V, \mbb{E} = \{E_0,E_1,\ldots,E_m \subseteq (V \times V)\}, \lambda)$, where $V$ is a set of vertices, $\mbb{E}$ is a family of edge sets, and $\lambda: E_k \rar \la [0,1], [0,1] \ra$ maps each edge in $E_k : k \leq m$ to an evidence tuple. The algebraic formulation of the presented evidential path algebra operates on an $n \times n \times m$ tensor representation of this network \cite{tensor:kolda2005}\footnote{The tensor representation can also be thought of as a set of $m$ adjacency matrix ``slices", where each matrix has its own label.}. The evidence tensor $\mca{A}$ is defined as
\begin{equation*}
	\mca{A}^{k}_{i,j} = 
		\begin{cases}
			\lambda(i,j) & \text{if } (i,j) \in E_k \\
			\la 0, 0 \ra & \text{otherwise}.
		\end{cases}
\end{equation*}
The two $n$ dimensions represent the vertices and the single $m$ dimension represents the various edge labels. Thus, there is a one-to-one mapping between a multi-relational, evidence network and an evidence tensor. The entries of the tensor denote the amount of positive ($w^+$) and negative ($w^-$) evidence for the edge $(i,j) \in E_k$, where $\la 0, 0 \ra$ denotes no evidence when no such edge exists. Inference on this tensor can be accomplished through the two binary operations
\begin{equation*}
	+:  \la [0,1], [0,1] \ra \times \la [0,1], [0,1] \ra \rar \la [0,1], [0,1] \ra
\end{equation*}
and
\begin{equation*}
	\cdot: \la [0,1], [0,1] \ra \times \la [0,1], [0,1] \ra \rar  \la [0,1], [0,1] \ra.
\end{equation*}
The function rules of these operations are
\begin{equation*}
	\la w_1^+, w_1^- \ra + \la w_2^+, w_2^- \ra = \la (w_1^+ + w_2^+), (w_1^- + w_2^-) \ra
\end{equation*}
and
\begin{footnotesize}
\begin{equation*}
	\la w_1^+, w_1^- \ra \cdot \la w_2^+, w_2^- \ra = \la (w_1^+ \cdot w_2^+), (w_1^+ \cdot w_2^- + w_1^- \cdot w_2^+ + w_1^- \cdot w_2^-) \ra.
\end{equation*}
\end{footnotesize}
These two operations form an algebraic ring over the evidence tensor (i.e.~multi-relational, evidence network). The binary operation $+$ is associative, has an identity of $\la 0,0 \ra$, and is commutative. The binary operation $\cdot$ is associative with an identity of $\la 1,0 \ra$. The operation $+$ supports the notion that evidence (from two independent experiences/inferences) can be summed together \cite{inherit:wang1994}. The operation $\cdot$ supports the notion that positive evidence can be multiplied to form new positive evidence and that conflicting and negative evidence accounts for negative evidence.

The next subsections will formalize the various syllogisms on inheritance edges in a multi-relational network and then the following subsection will discuss the application of these operations to any arbitrary path through an evidence tensor. Note that what is presented is a set of operations that will operate on $n \times n$ matrix ``slices" of the evidence tensor. In this respect, the presented operations are ``global" computations and thus, are computationally inefficient and contrary to AIKIR. However, these operations can be implemented as ``local" computations using various methods such grammar walks \cite{grammar:rodriguez2007} in specific areas of the network because evidential reasoning does not require a network-wide computation.

\subsection{Inferring Inheritance Evidence}

\begin{figure}[h!]	
	\begin{center}
		\includegraphics[width=0.32\textwidth]{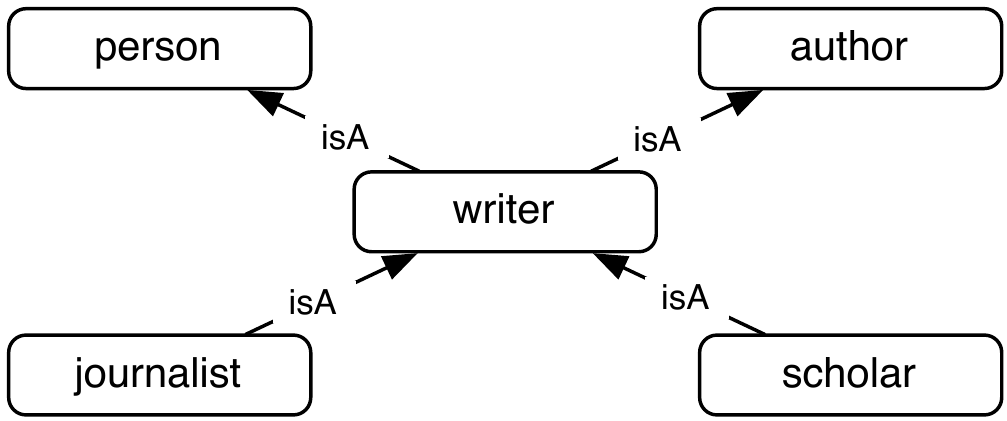}	
		\caption{\label{fig:example-network} An an inheritance network.}
	\end{center}
\end{figure}

Inheritance relations are handled as a special case situaion when reasoning in a multi-relational network. Example inheritance predicates in an RDF network are $\small{\ttt{rdf:type}}$ and $\small{\ttt{rdfs:subClassOf}}$. Figure \ref{fig:example-network} presents a simple example network that will be used to demonstrate the inference rules of deduction, induction, abduction, and exemplification.

\subsubsection{Deductive Inheritance}

Deduction is defined as a two step ``walk" on the inheritance component of an evidential network. A two step walk can be computed by squaring a matrix. Suppose a standard square $\{0,1\}$-matrix denoted $\mbf{A}$, where $\mbf{A}_{i,j} = 1$ if there is an edge between vertex $i$ and $j$, and $0$ otherwise. The $2^\text{nd}$ power of this matrix, as defined by ordinary matrix multiplication, $\mbf{A}\mbf{A}$, will yield a new matrix where entry $(\mbf{A}\mbf{A})_{i,j}$ denotes the total number of paths of length $2$ starting from vertex $i$ and ending on vertex $j$ \cite{graph:chartrand1977}. With respect to an evidence tensor, determining the product of two adjacency matrix ``slices", will yield a new adjacency matrix where the entry $(i,j)$ denotes the total amount of deductive evidence supporting $(i,j)$.

Evidential matrix multiplication is defined as ordinary matrix multiplication, but respective of the rules of $\cdot$ and $+$. Thus,
\begin{equation*}
	\left( \mca{A}^k\mca{A}^{k'} \right)_{i,j} =  \sum_{l \in V} \mca{A}_{i,l}^{k} \cdot \mca{A}_{l,j}^{k'} : k, k' \leq m .
\end{equation*}
In the degenerate case where all positive evidence is $1$ and all negative evidence is $0$, such that
\begin{equation*}
	\mca{A}^{k}_{i,j} = 
		\begin{cases}
			\la 1, 0 \ra & \text{if } (i,j) \in E_k \\
			\la 0, 0 \ra & \text{otherwise},
		\end{cases}
\end{equation*}
evidential matrix multiplication will set $w^+$ to the total number of paths from vertex $i$ to vertex $j$  and $0$ to $w^-$. In this form, the evidential path algebra yields results that are isomorphic to the original formulation of the path algebra in \cite{pathalg:rodriguez2008}.

\begin{figure}[h!]	
	\begin{center}
		\includegraphics[width=0.34\textwidth]{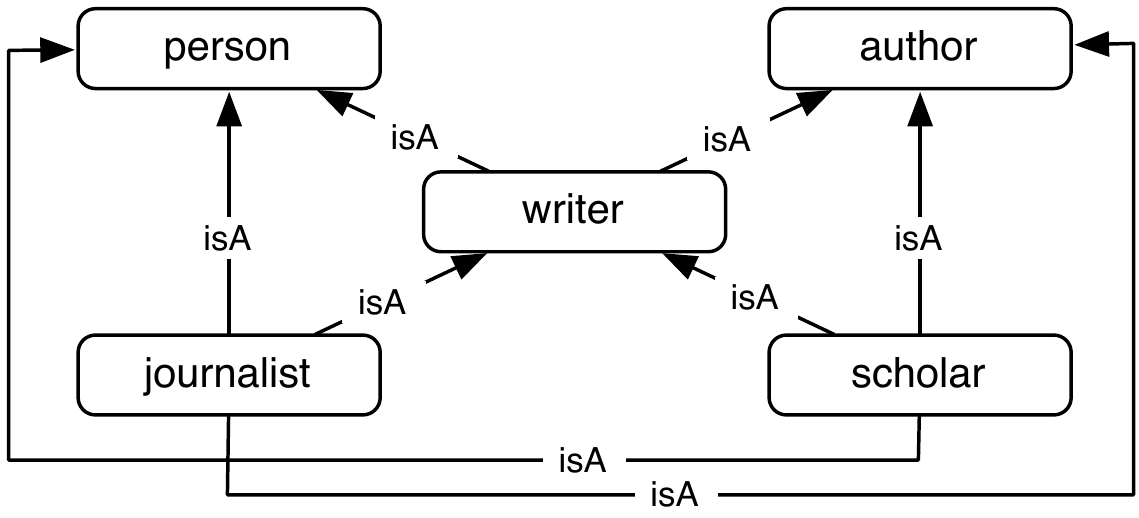}	
		\caption{\label{fig:deduced-network} Deduction in an inheritance network.}
	\end{center}
\end{figure}

In Figure \ref{fig:deduced-network}, deduction, as defined by $\mca{A}^{\small{\ttt{isA}}}\mca{A}^{\small{\ttt{isA}}}$, infers four new edges. Note that the evidence tuples are not presented in order to preserve diagram clarity.

\subsubsection{Inductive Inheritance}

Induction is the process of generalization given instances. In order to compute induction in an inheritance region of an evidence tensor, it is important to take the \textit{converse transpose} of an adjacency matrix ``slice". The operation of taking a statement like ``a scholar is a writer" and reversing it to derive the statement ``a writer is a scholar" is known as taking the converse of the statement. With respect to determining the evidence for the converse of a statement, all positive evidence for a ``scholar is a writer" is considered positive evidence that a ``writer is a scholar". However, all negative evidence for ``a scholar is a writer" should not be considered negative evidence for a ``writer is a scholar" as there is no evidence in the original statement for writers not being scholars. Such statement converses can be expressed using the evidential algebra. For standard matrices, the transpose of a matrix is defined as $(\mbf{A}^\top)_{i,j} = \mbf{A}_{j,i}$ and denotes reversing the direction of an edge (i.e.~taking the converse of a statement). However, for evidential edges, the converse transpose of an evidential matrix is defined as
\begin{equation*}
	\hat{\mca{A}}^k_{i,j} = \la \gamma^+(\mca{A}^k_{j,i}), 0 \ra,
\end{equation*}
where $\gamma^+: \la [0,1], [0,1] \ra \rar [0,1]$ maps an evidence tuple to its first component (i.e.~$w^+$ positive evidence). This operation ensures that the converse of an evidence tuple maintains no negative evidence.

\begin{figure}[h!]	
	\begin{center}
		\includegraphics[width=0.315\textwidth]{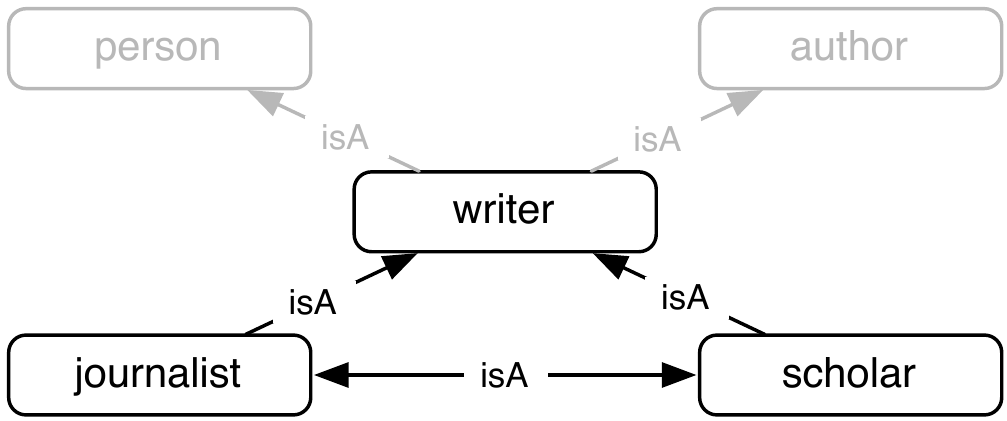}	
		\caption{\label{fig:induced-network} Induction in an inheritance network.}
	\end{center}
\end{figure}

In Figure \ref{fig:induced-network}, induction, as defined by $\mca{A}^{\small{\ttt{isA}}}\hat{\mca{A}}^{\small{\ttt{isA}}}$, infers two new ${\small{\ttt{isA}}}$ edges between $\small{\ttt{journalist}}$ and $\small{\ttt{scholar}}$. As stated previous, and to stress the point, the negative evidence for these two new evidence tuples is $0$.

\subsubsection{Abductive Inheritance}

\begin{figure}[h!]	
	\begin{center}
		\includegraphics[width=0.315\textwidth]{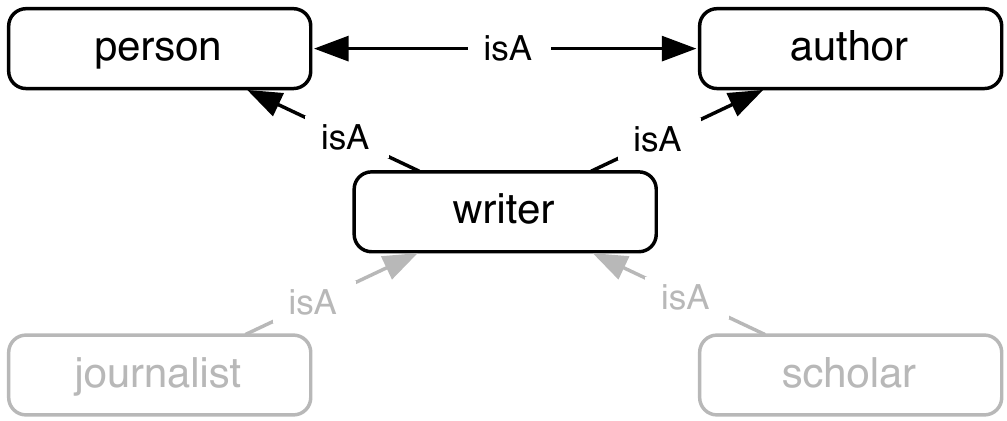}	
		\caption{\label{fig:abduced-network} Abduction in an inheritance network.}
	\end{center}
\end{figure}

Abduction is the reverse of induction. In Figure \ref{fig:abduced-network}, abduction, as defined by $\hat{\mca{A}}^{\small{\ttt{isA}}}\mca{A}^{\small{\ttt{isA}}}$, infers two new ${\small{\ttt{isA}}}$ edges between $\small{\ttt{person}}$ and $\small{\ttt{author}}$. Similar to induction, a converse transpose will generate no negative evidence for these two new evidence tuples.

\subsubsection{Exemplative Inheritance}

\begin{figure}[h!]	
	\begin{center}
		\includegraphics[width=0.34\textwidth]{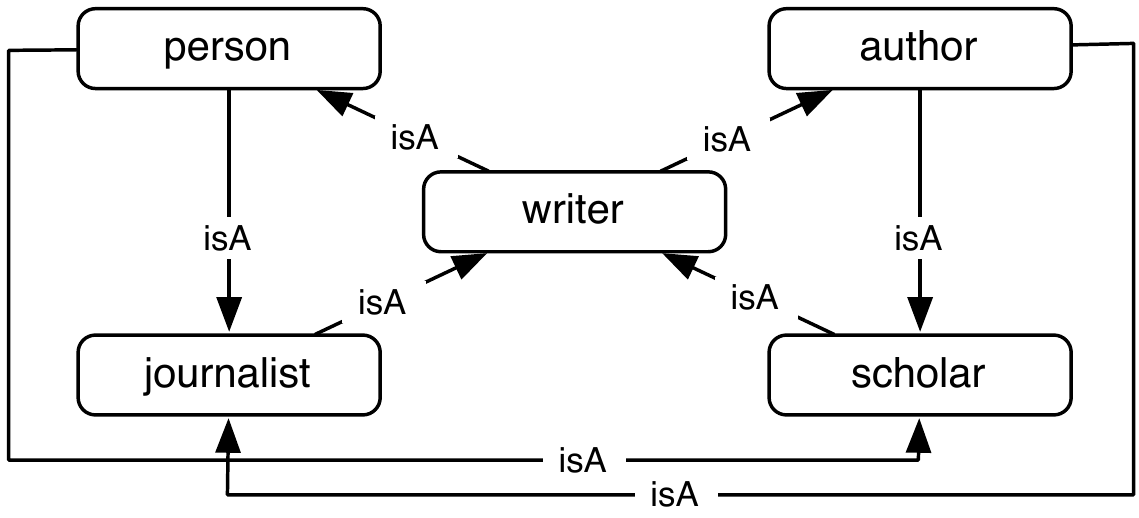}	
		\caption{\label{fig:exempuced-network} Exemplification in an inheritance network.}
	\end{center}
\end{figure}

Exemplary inheritance paths can be determined by the multiplication of two converse transpose matrices. In Figure \ref{fig:exempuced-network}, exemplification, as defined by $\hat{\mca{A}}^{\small{\ttt{isA}}}\hat{\mca{A}}^{\small{\ttt{isA}}}$, infers four new evidence tuples.

This subsection presented the syllogisms of deduction, induction, abduction, and exemplification and their use in an inheritance region of an evidence tensor. Note that this region may account for more than a single $m$-dimension as many labels can have an similar meaning to $\small{\ttt{isA}}$ (e.g.~$\small{\ttt{similarTo}}$, $\small{\ttt{equivalentTo}}$, $\small{\ttt{implies}}$, etc.). The next section will discuss reasoning using arbitrary paths through a multi-relational evidence network and thus, for those paths that may not necessarily contain $\small{\ttt{isA}}$ edges.

\subsection{Inferring Non-Inheritance Evidence}

\begin{figure}[h!]	
	\begin{center}
		\includegraphics[width=0.315\textwidth]{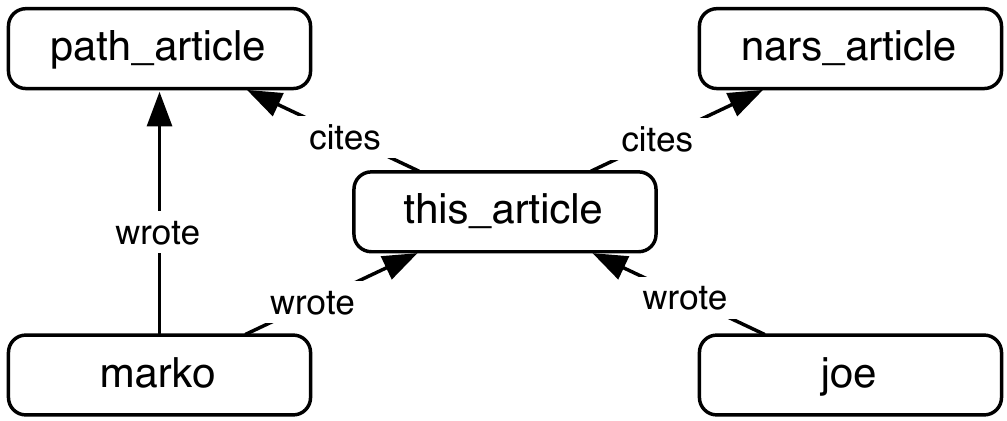}	
		\caption{\label{fig:arbitrary-network} A multi-relational knowledge network.}
	\end{center}
\end{figure}

A multi-relational network may be composed of various types of relationships. Figure \ref{fig:arbitrary-network} diagrams an example multi-relational network that will be referred to in the examples of this subsection\footnote{There is nothing that prevents the network in Figure \ref{fig:example-network} to be merged with the network in Figure \ref{fig:arbitrary-network} (e.g.~$\small{\ttt{marko}}$ and $\small{\ttt{joe}}$ are both $\small{\ttt{scholar}}$s). However, for diagram clarity, this is not represented.}.

The network in Figure \ref{fig:arbitrary-network} is composed of a reference to this article, the denoted authors of this article, and two citations from this article to other articles. In a bivalent logic, these statements are true because they exist. However, in scholarly publishing it is rare, nearly impossible, for two people to ``equally" write an article together. While ideas are shared and drafts are written, read, and edited, the article's final form is always a biased reflection of the approach of some authors over others. With respect to the statements diagrammed in Figure \ref{fig:arbitrary-network}, what is the evidence for these statements? The following descriptions are provided to expose, for each edge presented above, how much supporting or detracting evidence there is for $i$'s $m$-type relationship to $j$:
\newline
\begin{footnotesize}
\noindent $(\small{\ttt{marko}}, \small{\ttt{wrote}}, \small{\ttt{this\_article}})$:
	\begin{itemize}\setlength{\itemsep}{-4pt}
		\item $w^+$: notation, writing style, diagram style, american spelling
		\item $w^-$: logic, reasoning, citations, philosophy
	\end{itemize}
$(\small{\ttt{marko}}, \small{\ttt{wrote}}, \small{\ttt{path\_article}})$:
	\begin{itemize}\setlength{\itemsep}{-4pt}
		\item $w^+$: notation, writing style
		\item $w^-$: algebra, no diagrams
	\end{itemize}
 $(\small{\ttt{joe}}, \small{\ttt{wrote}}, \small{\ttt{this\_article}})$:
 	\begin{itemize}\setlength{\itemsep}{-4pt}
		\item $w^+$: logic, reasoning, citations, philosophy, rings
		\item $w^-:$ notation, writing style, diagram style, american spelling
	\end{itemize}
$(\small{\ttt{this\_article}}, \small{\ttt{cites}}, \small{\ttt{path\_article}})$:
 	\begin{itemize}\setlength{\itemsep}{-4pt}
		\item $w^+$: \cite{pathalg:rodriguez2008} citation, rings
		\item $w^-$: graph notation, philosophy, only a single algebra citation
	\end{itemize}
$(\small{\ttt{this\_article}}, \small{\ttt{cites}}, \small{\ttt{nars\_article}})$:
 	\begin{itemize}\setlength{\itemsep}{-4pt}
		\item $w^+$: \cite{nars2.2:wang} citation, evidential notation, syllogisms
		\item $w^-$: path expressions, semantic web, rdf, owl, rdfs
	\end{itemize}
\end{footnotesize}

The presented positive and negative evidence ``metadata" (e.g.~writing style, american spelling, citation patterns, etc.) can be represented in a multi-relational network. From this multi-relational encoding, it is possible, through automated means, to infer new evidence or revise existing evidence in the network with prescribed path expressions. In other words, a region of the network can provide further supporting and/or detracting evidence for another region of the network. In order to demonstrate two examples of this, the multi-relational network in Figure \ref{fig:arbitrary-network} will be used in this section to
 \begin{enumerate}\setlength{\itemsep}{-3pt}
	\item infer new independent evidence supporting the claims that $\small{\ttt{marko}}$ and $\small{\ttt{joe}}$ $\small{\ttt{wrote}}$ $\small{\ttt{this\_article}}$ according to the notion that self-citations are positive evidence supporting authorship, and
	\item infer new evidence that $\small{\ttt{marko}}$ and $\small{\ttt{joe}}$ have a $\small{\ttt{coauthor}}$ship edge between them.
\end{enumerate}

\subsubsection{Self-Citation Paths}

Suppose that the article citation evidence in Figure \ref{fig:arbitrary-network} was experienced by the agent (e.g.~a repository feed) and added to its existing internal network. At the point of insertion, it is possible to revise the evidence tuple for $\small{\ttt{wrote}}$ based upon the idea that self-citations in an article are considered evidence for authorship of that article \cite{metadata:rodriguez2006}. Furthermore, in order to ensure independent evidence, assume that the current evidence for $\small{\ttt{wrote}}$ in Figure \ref{fig:arbitrary-network} was not determined using self-citation information\footnote{Refer to \cite{nars2.2:wang} for the definition and importance of independent evidence in evidential logics.}. Given this scenario, the following inference rule will update all $\small{\ttt{wrote}}$ evidence according to inferred self-citation evidence:
\begin{equation*}
	\mca{A}^{\small{\ttt{wrote}}}_{(t+1)} = \left( \left( c(\mca{A}^{\small{\ttt{wrote}}}_{(t)}) \mca{A}^{\small{\ttt{cites}}}_{(t)} {\mca{A}^{\small{\ttt{wrote}}}_{(t)}}^\top \right) \circ \mbf{I} \right)+ \mca{A}^{\small{\ttt{wrote}}}_{(t)},
\end{equation*}
where $t \in \mbb{N}^+_0$ is the current time step, $c(\mca{A}^{\small{\ttt{wrote}}})$ ``clips" the evidence in $\mca{A}^{\small{\ttt{wrote}}}$, $\circ$ is the entry-wise Hadamard multiplication operation\footnote{For review, Hadamard entry-wise multiplication is defined as
\begin{scriptsize}
\begin{equation*}
	\mbf{A} \circ \mbf{B} = \left[
	\begin{array}{ccc}
		\mbf{A}_{1,1} \cdot \mbf{B}_{1,1} & \cdots & \mbf{A}_{1,j} \cdot \mbf{B}_{1,j} \\
		\vdots & \ddots & \vdots \\
		\mbf{A}_{i,1} \cdot \mbf{B}_{i,1} & \cdots & \mbf{A}_{i,j} \cdot \mbf{B}_{i,j} \\
	\end{array} \right ] .
\end{equation*}
\end{scriptsize}}, and $\mbf{I}$ is the evidential identity matrix 
\begin{equation*}
	\mbf{I}_{i,j} = 
		\begin{cases}
			\la 1, 0 \ra & \text{if } i = j \\
			\la 0, 0 \ra & \text{otherwise}.
		\end{cases}
\end{equation*}
In words, the self-citation inference rule states that evidence for $\mca{A}^{\small{\ttt{wrote}}}$ can be modulated by the total evidence for the path that goes from an author, to their written articles, to the articles that those articles cite, and then finally, to the authors of those cited articles. However, in order to ensure that those cited authors are the original author from the start of the path (i.e.~self-citation), it is important to filter on the identity matrix $\mbf{I}$. Hadamard entry-wise multiplication is used to apply a matrix filter to a path. Note that the transpose of an evidence matrix, not the converse transpose of the evidence matrix, is used when taking the converse of a non-inheritance statement. The reason for this is that the positive and negative evidence for the statement ``marko wrote this article" is the same for ``this article was written by marko". Next, inferred evidence for  $\small{\ttt{wrote}}$ must be independent of the evidence used to calculate it. Thus, $\mca{A}^{\small{\ttt{wrote}}}$ is mapped to a $(\la 1,0 \ra, \la 0, 0 \ra)$-matrix using the clip $c$ operation, where
\begin{equation*}
	c(\mca{A}^k)_{i,j} = 
		\begin{cases}
			\la 1,0 \ra & \text{if } \mca{A}^k_{i,j} \neq \la 0, 0 \ra \\
			\la 0, 0 \ra & \text{otherwise}.
		\end{cases}
\end{equation*}
Finally, the total evidence for the self-citation path is summed with the current evidence for the $\small{\ttt{wrote}}$ edge. Thus, the agent has used self-citations as further, revising evidence for $\small{\ttt{wrote}}$.

\subsubsection{Coauthorship Paths}

\begin{figure}[h!]	
	\begin{center}
		\includegraphics[width=0.315\textwidth]{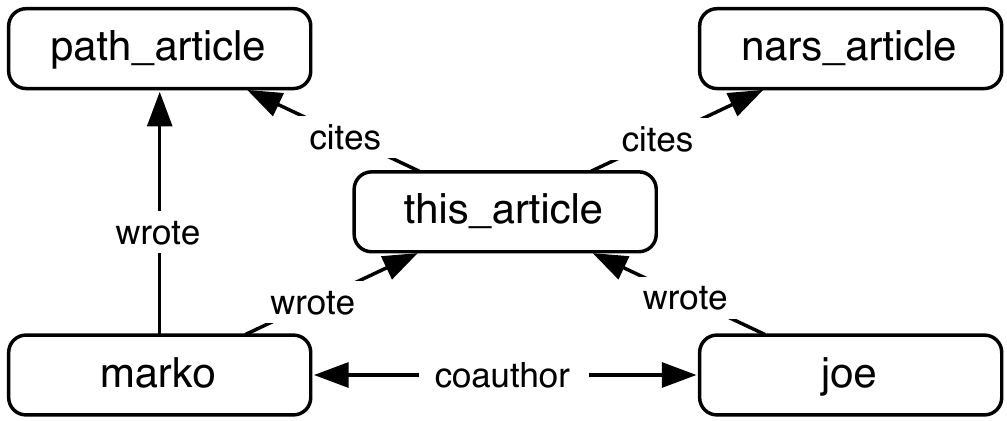}	
		\caption{\label{fig:coauthor-network} Coauthoring in an inheritance network.}
	\end{center}
\end{figure}

Two people are considered coauthors if they have both written an article together. The evidence for coauthorship is determined by the total evidence across all their jointly written articles. In its algebraic form, the evidence for $\small{\ttt{coauthor}}$ can be determined by
\begin{equation*}
	\mca{A}^{\small{\ttt{coauthor}}}_{(t+1)} = \left( \left( \mca{A}^{\small{\ttt{wrote}}}_{(t)} {\mca{A}^{\small{\ttt{wrote}}}_{(t)}}^\top \right) \circ n(\mbf{I}) \right) + \mca{A}^{\small{\ttt{coauthor}}}_{(t)},
\end{equation*}
where $n(\mbf{I})$ ``nots" the evidential identity matrix such that such that every $\la 1, 0 \ra$ is a $\la 0, 0 \ra$ and every $\la 0, 0 \ra$ is a $\la 1, 0 \ra$. The reason for the $n(\mbf{I})$ filter is that to represent a $\small{\ttt{coauthor}}$ path from a person to their authored papers and then to \textit{other} authors of those papers, the path must exclude the original author as an author is not a coauthor of themselves. In other words, it must filter out the identity evidence matrix. The two inferred $\small{\ttt{coauthor}}$ edges between $\small{\ttt{marko}}$ and $\small{\ttt{joe}}$ are diagrammed in Figure \ref{fig:coauthor-network}.

The article \cite{pathalg:rodriguez2008} provides an in-depth review of different inferences that can be made with arbitrary paths, various filters, and how the theorems of the general path algebra can be applied to derive equivalent, yet more computationally efficient paths. The examples presented in \cite{pathalg:rodriguez2008} can be applied to an evidence tensor as long as the definitions of $+$ and $\cdot$, as defined in this article, are respected.

\section{Conclusion}

Reasoning with axiomatic logics is computationally expensive \cite{donini:complexreason2002,reasoning:fensel2007}. With respect to the Semantic Web, and with the integration capabilities brought forth by the Linked Data initiative, such reasoning is intractable. The original assumption driving the development of NARS is the Assumption of Insufficient Knowledge and Insufficient Resources (AIKIR) \cite{nars2.2:wang}. Given space and time constraints, an agent cannot reason over the entire Semantic Web, and potentially, not even over its entire internal knowledge network. The benefit of the inheritance-based syllogisms are that they do not require a global analysis of the knowledge network, can be executed independently of each other, and at the their core, are very simple and computationally efficient. With respect to inferencing with arbitrary path expressions, the efficiency is dependent on the length of the path and the number of applied filters. It is important to note that the matrix formalism presented in this article is very much intractable as the best known algorithm to compute ordinary matrix multiplication is approximately $O(|V|^{2.807})$ \cite{matrix:strassen1969}. As stated previously, this matrix model can be approximated using various techniques such as grammar walk algorithms which do not compute the inferences over the entire network, but instead, on local subgraphs (i.e.~paths starting from particular vertices) \cite{grammar:rodriguez2007}. With evidential logic, such walks can be executed when resources are available and only in those areas of the knowledge network where it is deemed necessary (e.g.~$f \sim 0.5$ and/or low $c$ areas).

Finally, to actually represent a multi-relational, evidence network in RDF and on the Semantic Web, some form of reification can be used. A popular technique is the quad-form of a ``triple" where a statement maintains a fourth component known as a named graph \cite{named:carroll2005}. With reification, statements can be attached to statements and thus, a $\la w^+, w^- \ra$ evidence tuple can be assigned to an RDF statement.

This article presented a non-axiomatic evidential logic that can be implemented within the constructs of RDF and thus, can be used as an evidential reasoning system in the Semantic Web. The benefit of this system is that it works with arbitrary, partial, and contradictory knowledge while, at the same time, in a non-matrix implementation, supports a tractable approximate reasoning process.

\section{Acknowledgements}

Vadas Gintautas provided useful insight during the development of these ideas. This work was funded by a Director's Fellowship from the Los Alamos National Laboratory, EPSRC grant EP/D504376/1, and British Telecom plc.

\end{document}